\documentclass[useAMS,usenatbib,12pt,epsfig]{article}
\usepackage[X2,T2A]{fontenc}
\usepackage[koi8-r]{inputenc}
\usepackage[english]{babel}
\usepackage{amssymb}
\usepackage{graphicx}
\usepackage{cite}
\usepackage{wrapfig}
\usepackage{epsf}

\begin{document}
\selectlanguage{english}

\thispagestyle{empty}

\bigskip
\bigskip
\bigskip
\begin{center}
\begin{Large}
{Effect of Radiative Feedbacks for Resonant Transitions 
during Cosmological Recombination}
\end{Large}
\end{center}
\bigskip
\bigskip
\bigskip
\begin{center}
{Kholupenko E.E.$^1$, Ivanchik A.V.$^{\rm 1,2}$, Varshalovich D.A.$^{\rm 1,2}$}
\end{center}
{$^{\rm 1}$Ioffe Physical-Technical Institute, St.-Petersburg 194021, Russia}
\\{$^{\rm 2}$ Cosmic Research Chair of the St.-Petersburg State Polytechnical 
University, Russia}
\bigskip
\bigskip
\bigskip
\bigskip
\begin{center}
\bf{Abstract}
\end{center}
{The inhibition of the total HI $n\leftrightarrow 1$ transition rate by 
delayed resonant reabsorption of HI $(n+1)\rightarrow 1$ photons by 
HI $n\rightarrow 1$ line which is possible due to cosmological redshift 
is considered semi-analytically. 
The method taking into account this effect in the frame of simple 
three-level approximation model of recombination is suggested. 
It is confirmed that the resonant feedbacks affect ionization fraction 
at the level about 0.2\% for the epoch of last scattering. 

Similar consideration of HeI $2^1P\leftrightarrow 1^1S$ $\Rightarrow$ 
HeI $2^3P\leftrightarrow 1^1S$ feedback for helium is provided. 
It is confirmed that allowance of this feedback leads to 
increase of predicted free electron fraction by 0.12\% at $z\simeq 2300$. 
It is shown that taking into account absorption and thermalization of 
HeI $2^1P\leftrightarrow 1^1S$ resonant superequilibrium photons 
(during their redshifting to the HeI $2^3P\leftrightarrow 1^1S$ frequency) 
by small amount of neutral hydrogen ($10^{\rm -7} - 10^{\rm -4}$ of total number of 
hydrogen atoms and ions) existing in helium recombination epoch is 
important for correct consideration of this helium feedback.
}

\newpage
\section{Introduction}
\hspace{1.1cm}
Cosmological recombination plays an important role for the observations  
of the CMBR anisotropy, because the visibility of epoch at fixed $z$ 
directly depends on concentration of free electrons. Today it is clear 
that for correct analysis of experimental data from Planck mission 
(launched at May 2009) and other future experiments on anisotropy 
measurements the ionization fraction as a function of redshift should be 
calculated with relative accuracy within 0.1\% at least. Also the codes for 
such calculations should be fast enough for effective determination of 
cosmological parameters from CMBR anisotropy data. Several approaches 
are suggested to satisfy these conditions: three-level approximation (TLA) with 
fudge factors (e.g. recfast, Seager et al. 1999, Wong et al. 2008), 
training-conception (e.g. RICO, Fendt et al. 2009), multy-level 
codes with not very large principal quantum number 
($5 \lesssim n_{\rm max}\lesssim 10$), and others. In a number of works 
the different approach was described (e.g. Dubrovich and Grachev 2005) and 
used (e.g. Kholupenko and Ivanchik 2006): 
the TLA-model has been successfully modified to include transitions 
from high excited states, helium recombination through ortho-channel, 
HI Ly$\alpha$ $\Rightarrow$  HI 2s$\leftrightarrow$1s feedback, 
and hydrogen continuum absorption. 
These works show that the way of modification of 
TLA-model appears fruitful and in principal in this way TLA-model can be 
developed to the form taking into account cumulative action of ``thin'' 
effects considered until now for precise calculations of cosmological 
recombination. 

The radiative feedbacks for the cosmological recombination problem 
have been considered in a number of works (Chluba and Sunyaev 2007, 2009; 
Switzer and Hirata 2008). 
The method presented in this work is similar to the method 
developed by Chluba and Sunyaev (2009), but in present work 
some additional analytical approximations have been obtained. 
Use of these approximations allows us to speed up 
the recombination calculations.
The main purpose of this paper is to show how the effect of 
HI Ly$(n+1)$ $\Rightarrow$  HI n$\leftrightarrow$1 
feedbacks for hydrogen (Chluba and Sunyaev 2007, 2009) and 
HeI $2^1P\leftrightarrow 1^1S$ $\Rightarrow$  HeI $2^3P\leftrightarrow 1^1S$ 
feedback for helium (Switzer and Hirata 2008, Chluba and Sunyaev 2009) 
can be taken into account in the frame 
of TLA-model and to provide an independent calculation of these effects for 
comparison. 

\section{Cosmological model}
All calculations have been performed in the frame of 
standard cosmological model. Corresponding values of cosmological 
parameters are indicated in Tab. 1. 
Dependence of Hubble constant on redshift is given by the following 
\begin{equation}
H(z) = H_0 \sqrt{\Omega_\Lambda+\Omega_{\rm m} (1+z)^3 + \Omega_{\rm rel}(1+z)^4}
\label{Hubble_const}
\end{equation}
Total concentration of atoms and ions for the certain component of plasma 
depends on redshift by power law: $N\sim (1+z)^3$.
\begin{table}
\begin{center}
\caption{
  Parameters of the standard cosmological model
  }
\begin{tabular}{lll}
  \hline
  Value description & Symbol & Value \\
  \hline
  total matter & $\Omega_{\rm tot}$ & ~~~~1 \\
  (in the units of critical density) & & \\
%  space curvature& $\Omega_k$ & ~~~~0 \\
  non-relativistic matter& $\Omega_{\rm m}=\Omega_{\rm CDM}+\Omega_{\rm b}$ & $0.27$ \\
  baryonic matter& $\Omega_{\rm b}$ & $0.045$ \\
  relativistic matter& $\Omega_{\rm rel}=\Omega_{\rm \gamma}+\Omega_{\rm \nu}$ & $ 8.23\cdot 10^{\rm -5}$ \\
  vacuum-like energy& $\Omega_\Lambda$ & $0.73$\\
  Hubble constant & $H_0$ & 70 km/s/Mpc \\
  radiation temperature & $T_{\rm 0}$ & $2.725$ K \\
  helium mass fraction & $Y$ & $0.24$ \\
  deuterium-to-hydrogen ratio  & $[D/H]$ & $3\cdot 10^{\rm -5}$ \\
  \hline
\end{tabular}
\end{center}
\end{table}

\section{Hydrogen Feedbacks}
\subsection{Semi-analytical Consideration of Resonant Feedbacks for Hydrogen}
The $n\leftrightarrow 1$ transition rate is given by the following formula
\begin{equation}
J_{\rm n}\simeq A_{\rm n}\left(N_{\rm n}\left(1+\eta(\nu_{\rm n})\right)
-{g_{\rm n}\over g_{\rm 1}}\eta(\nu_{\rm n})N_{\rm 1}\right)
\label{rate_definition}
\end{equation}
where $A_{\rm n}$ [s$^{\rm -1}$] is the coefficient of spontaneous radiative 
$n\rightarrow 1$ transition, $N_{\rm n}$ [cm$^{\rm -3}$] is the concentration of 
atoms in the state $n$, $N_{\rm 1}$ [cm$^{\rm -3}$] is the concentration of 
atoms in the ground state, $\eta(\nu)$ is the number of photons per 
$\nu$ mode, $\nu_{\rm n}$ is the central frequency of $n\rightarrow 1$ line,
$g_{\rm n}$, $g_{\rm 1}$ are the statistical weights of state $n$ and the ground state. 

\begin{wrapfigure}[34]{o}{0.9\textwidth}
\begin{center}
\includegraphics[width=0.9\textwidth, clip]{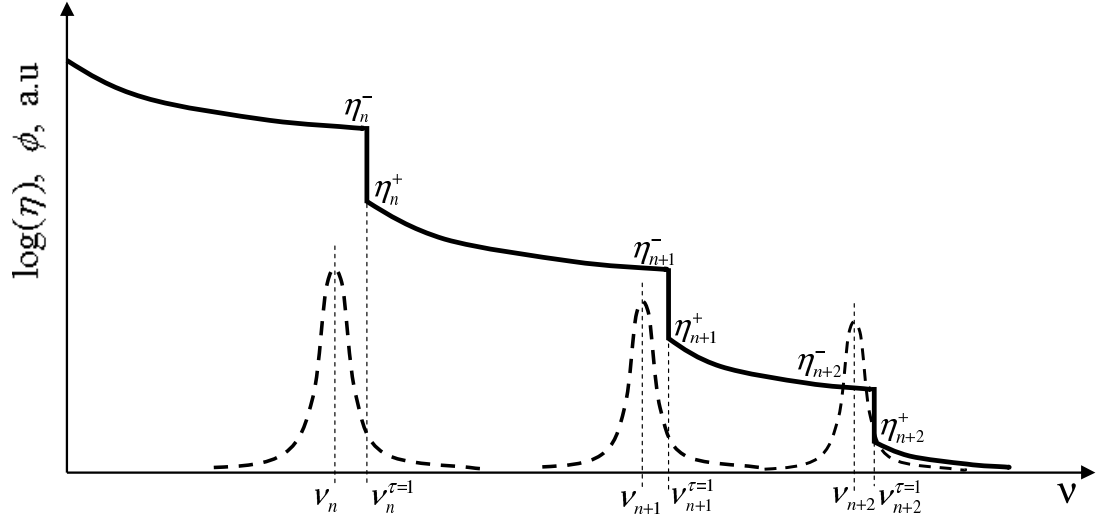}
\end{center}
\caption{Schematic picture of superequilibrium radiation transfer: 
occupation number $\eta$ as a function of frequency $\nu$ is shown by 
thick solid line, absorption line profiles $\phi$ as functions of frequency 
$\nu$ are shown by thick dashed line. Value $\nu_{n}$ is the central frequency 
of $n\rightarrow 1$ line profile, value $\nu^{\tau=1}_{n}$ is the frequency at
which the optical depth in $n\rightarrow 1$ line achieves unity.}
\label{fig1}
\end{wrapfigure}

Use of Sobolev solution allows us to rewrite (\ref{rate_definition}) 
in more convenient form:
\begin{equation}
J_{\rm n}\simeq P_{\rm n}A_{\rm n}\left(N_{\rm n}\left(1+\eta^{+}_{\rm n}\right)
-{g_{\rm n}\over g_{\rm 1}}\eta^{+}_{\rm n}N_{\rm 1}\right)
\label{Sobolev_rate_definition}
\end{equation}
where $P_{\rm n}$ is the usual Sobolev escape probability, $\eta^{+}_{\rm n}$ 
is the occupation number of radiation in the blue wing of $n\rightarrow 1$ 
line (see Fig. \ref{fig1}). 

Let us define the coefficient $C_n$ (the factor by 
which ``ordinary'' Sobolev transition rate (corresponding to the presence 
of proper resonant radiation only) is inhibited by the feedback effect):
\begin{equation}
C_{\rm n}=J_{\rm n}/J^{\rm 0}_{\rm n}
\end{equation}
where $J^{\rm 0}_{\rm n}$ is the $n\leftrightarrow 1$ transition rate in the case 
of equilibrium radiation in the blue wing of $n\leftrightarrow 1$ line 
(i.e. $\eta^{+}_{\rm n}=\eta^{\rm 0}_{\rm n}$ where $\eta^{\rm 0}_{\rm n}$ is equilibrium 
(planckian) occupation number at frequency $\nu_{\rm n}$). Neglecting induced 
$n\rightarrow 1$ transitions (in comparison with spontaneous $n\rightarrow 1$ 
transitions, since $\eta^{+}_{\rm n}\lesssim 10^{\rm -8}\ll 1$ for any reasonable case) 
one can obtain: 
\begin{equation}
C_{\rm n}={N_{\rm n}-\left(g_{\rm n}/g_{\rm 1}\right)\eta^{+}_{\rm n}N_{\rm 1} \over 
N_{\rm n}-\left(g_{\rm n} / g_{\rm 1}\right)\eta^{\rm 0}_{\rm n}N_{\rm 1}}
\end{equation}
Using quasistacionary solution of radiation transfer problem for 
$n\rightarrow 1$ line:
\begin{equation}
\eta={g_{\rm 1}N_{\rm n}\over g_{\rm n}N_{\rm 1}}\left(1-e^{\rm -\tau}\right)
+\eta^{+}e^{\rm -\tau}
\label{radiation_transfer_solution}
\end{equation}
and taking into account that for all Ly-lines optical depth $\tau$ is much 
larger than unity at the central frequency and in the red wing, one can write: 
\begin{equation}
\eta_{\rm n}\simeq\eta^{-}_{\rm n}\simeq 
{g_{\rm 1}N_{\rm n} \over g_{\rm n}N_{\rm 1}}
\label{quasistationary_population}
\end{equation} 
where $\eta^{-}_{\rm n}$ is the occupation number of radiation in the 
red wing of $n\rightarrow 1$ line (see Fig. \ref{fig1}). Using 
(\ref{quasistationary_population}) 
one can obtain the following expression for $C_{\rm n}$:
\begin{equation}
C_{\rm n}={\eta^{-}_{\rm n}-\eta^{+}_{\rm n} \over 
\eta^{-}_{\rm n} - \eta^{\rm 0}_{\rm n}}
\label{C_n_eta}
\end{equation}

In accordance with Zeldovich et al. (1968) and Peebles (1968) 
the populations of excited states are approximately determined 
relative to the state $n=2$ by the Boltzmann distribution:
\begin{equation}
{N_{\rm n}\over N_{\rm 2}}={g_{\rm n}\over g_{\rm 2}}
\exp\left(-{E_{\rm n}-E_{\rm 2}\over k_{\rm B}T}\right)
\label{Boltzmann_distrib}
\end{equation}
Using (\ref{quasistationary_population}) and (\ref{Boltzmann_distrib}) it is 
easy to show (e.g. Peebles 1968) that:
\begin{equation}
\eta^{-}_{\rm n}=\eta^{-}_{\rm 2}\exp\left(-{E_{\rm n}-E_{\rm 2}\over k_{\rm B}T}\right)
\end{equation}
or approximately 
\begin{equation}
\eta^{-}_{\rm n}=\eta_{\rm \alpha}{\eta^{\rm 0}_{\rm n} \over \eta^{\rm 0}_{\rm \alpha}}
\label{eta_n_from_alpha}
\end{equation}
where $\eta_{\rm \alpha}=\eta^{-}_{\rm 2}$ is the HI Ly$\alpha$ occupation number 
(additional subscript $\alpha$ emphasizes the important role of 
HI Ly$\alpha$ radiation for consideration of cosmological recombination). 

It is well known that in the absence of absorption the value $\eta^{+}_{\rm n}$ 
can be found by using the following formula:
\begin{equation}
\eta^{+}_{\rm n}(z)=\eta^{-}_{\rm n+1}(z'_{\rm n})
\label{eta_n_from_eta_n1}
\end{equation}
where $z'_{\rm n}$ for this case is given by the following relation
\begin{equation}
z'_{\rm n}=(1+z){\nu_{\rm n+1}\over \nu_{\rm n}}-1=(1+z){1-(n+1)^{\rm -2}\over 1-n^{\rm -2}}-1
\label{z_n}
\end{equation}
Using (\ref{eta_n_from_alpha}) one can obtain from (\ref{eta_n_from_eta_n1}) 
the following:
\begin{equation}
\eta^{+}_{\rm n}(z)=\eta^{\rm 0}_{\rm n+1}(z'_{\rm n})
{\eta_{\rm \alpha}(z'_{\rm n}) \over \eta^{\rm 0}_{\rm \alpha}(z'_{\rm n})}
\label{eta_minus_from_alpha}
\end{equation}
Substituting (\ref{eta_n_from_alpha}) and (\ref{eta_minus_from_alpha}) 
into (\ref{C_n_eta}) one can find that 
\begin{equation}
C_{\rm n}={\left(\eta_{\rm \alpha} / \eta^{\rm 0}_{\rm \alpha}\right)
-\left(\eta^{\rm 0}_{\rm n+1}(z'_{\rm n}) / \eta^{\rm 0}_{\rm n}(z)\right)
\left(\eta_{\rm \alpha}(z'_{\rm n}) / \eta^{\rm 0}_{\rm \alpha}(z'_{\rm n})\right)
\over \left(\eta_{\rm \alpha} / \eta^{\rm 0}_{\rm \alpha}\right) - 1}
\end{equation}
Note that definition (\ref{z_n}) results in 
$\eta^{\rm 0}_{\rm n+1}(z'_{\rm n}) = \eta^{\rm 0}_{\rm n}(z)$. 
Using this one can obtain the following formula for $C_{\rm n}$:
\begin{equation}
C_{\rm n}={\left(\eta_{\rm \alpha} / \eta^{\rm 0}_{\rm \alpha}\right)
-\left(\eta_{\rm \alpha}(z'_{\rm n}) / \eta^{\rm 0}_{\rm \alpha}(z'_{\rm n})\right)
\over \left(\eta_{\rm \alpha} / \eta^{\rm 0}_{\rm \alpha}\right) - 1}
\end{equation}

\begin{wrapfigure}[34]{o}{0.9\textwidth}
\begin{center}
\includegraphics[width=0.9\textwidth, clip]{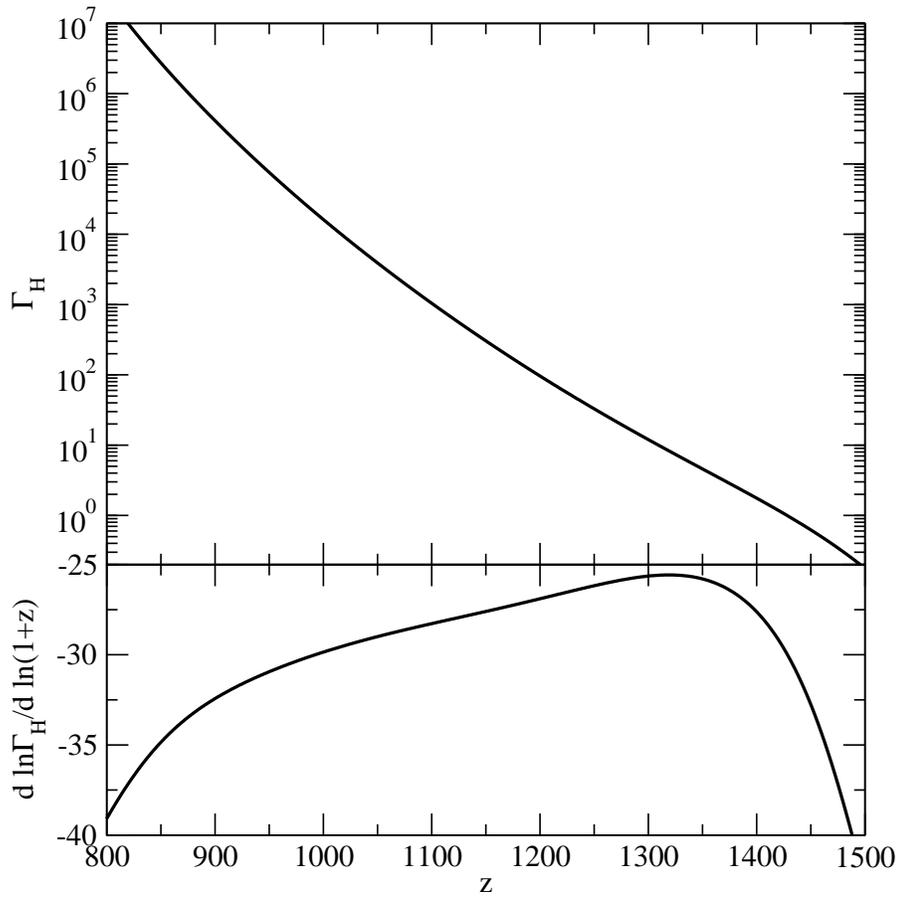}
\end{center}
\caption{
{\bf Top panel:} Dependence of $\Gamma_{\rm H}$ on redshift $z$. 
{\bf Bottom panel:} Dependence of $d\ln\Gamma_{\rm H}/d\ln\left(1+z\right)$ 
on redshift $z$.}
\label{fig2}
\end{wrapfigure}

Let us define the following function:
\begin{equation}
\Gamma_{\rm H}={\eta_{\rm \alpha} \over \eta^{\rm 0}_{\rm \alpha}} - 1
\label{Gamma_H}
\end{equation}
which has sense of relative overheating of Ly$\alpha$ 
radiation in comparison with its equilibrium value (see top panel of 
Fig. \ref{fig2}). Using this function the formula for $C_{\rm n}$ can be 
rewritten in the following form:
\begin{equation}
C_{\rm n}=1-\Gamma_{\rm H}(z'_{\rm n})/\Gamma_{\rm H}(z)
\label{C_n_main}
\end{equation}

Formula (\ref{C_n_main}) shows us that in the first approximation 
the knowledge of the {\bf only} function 
$\Gamma_{\rm H}(z)$ is enough for calculating {\bf all} the feedback 
inhibition coefficients $C_{\rm n}$. 

Asymptotic of $C_{\rm n}$ at large $n$ can be obtained by using Taylor's 
expansion for $\Gamma_{\rm H}(z'_{\rm n})$:
\begin{equation}
\Gamma_{\rm H}(z'_{\rm n})\simeq \Gamma_{\rm H}(z)+{d\Gamma_{\rm H}(z) \over dz}
\left(z'_{\rm n}-z\right)
\label{Taylor1}
\end{equation}
For the large $n$ the following approximation is valid 
(e.g. Grin and Hirata 2009)
\begin{equation}
z'_{\rm n}\simeq z+2(1+z)n^{\rm -3}
\label{z_approx}
\end{equation}
Using (\ref{Taylor1}) and (\ref{z_approx}) one can obtain
\begin{equation}
C_{\rm n}\simeq -{2(1+z) \over n^3}{d\ln \Gamma_{\rm H} \over dz}=
-{2 \over n^3}{d\ln \Gamma_{\rm H} \over d\ln\left(1+z\right)}
\label{assympt_C}
\end{equation}
Formula (\ref{assympt_C}) is valid for principal quantum numbers $n\ge 16$ 
and redshifts $z = 800 - 1800$ with relative accuracy within 10\%. 
Dependence of ${d\ln \Gamma_{\rm H} / d\ln\left(1+z\right)}$ on $z$ 
is shown in the bottom panel of Fig. \ref{fig2}. Note that this 
dependence is almost linear for $z= 1000 - 1300$. Taking this into 
account one can find simple approximation of $\Gamma_{\rm H}$ for 
different values of the key cosmological parameters. Such an approximation 
may be useful for fast calculations of full set of inhibition coefficients 
$C_{n}$ without preliminary calculation of $\Gamma_{\rm H}$.

Formula (\ref{assympt_C}) shows that at any fixed $z$ values 
$C_n$ decrease with increasing $n$ by power law $n^{\rm -3}$. This 
means that beginning from some value $n_{\rm max}$ the total influence 
of transitions from high excited states with $n\ge n_{\rm max}$ and 
feedback effect becomes negligible (at any given level of accuracy).

It should be noted that $C_n$ are the functions of non-equilibrium 
radiation field at different moments $z'$ (shifted relative current moment $z$). 
This radiation field depends on solution of cosmological recombination 
problem which in turn depends on $C_n$ in general case. 
This means that for maximal accuracy of numerical solution of 
considered problem one should use iteration method 
(Chluba and Sunyaev 2007, Switzer and Hirata 2008). 
Nevertheless it is evident that influence of taking into account 
$C_{\rm n}$ on ionization history  
and non-equilibrium radiation field is small, i.e. correction 
to intensity of non-equilibrium radiation is much less than unity 
(no more than several percent). This allows us to use unperturbed (i.e. 
calculated in the frame of standard TLA model without feedbacks) value of 
Ly$\alpha$ radiation field for determination of function $\Gamma_H(z)$. 

Coefficients $C_n$ may be useful not only for improvement of TLA model 
but also for using in multilevel codes (without splitting on angular moment 
(Grachev and Dubrovich 1991) and with splitting on angular moment 
(Burgin 2003, Rubino-Martin et al. 2006, Switzer and Hirata 2008, Grin and Hirata 2009)) 
for precise calculations of 
ionization history and recombination distortions of CMBR spectrum. 
As well as formalism of Sobolev escape probability,  
the formalism of inhibition coefficients $C_n$ allows us to avoid 
direct solution of radiation transfer equations for lines and 
transitions under consideration. 

\begin{wrapfigure}[34]{o}{0.9\textwidth}
\begin{center}
\includegraphics[width=0.9\textwidth, clip]{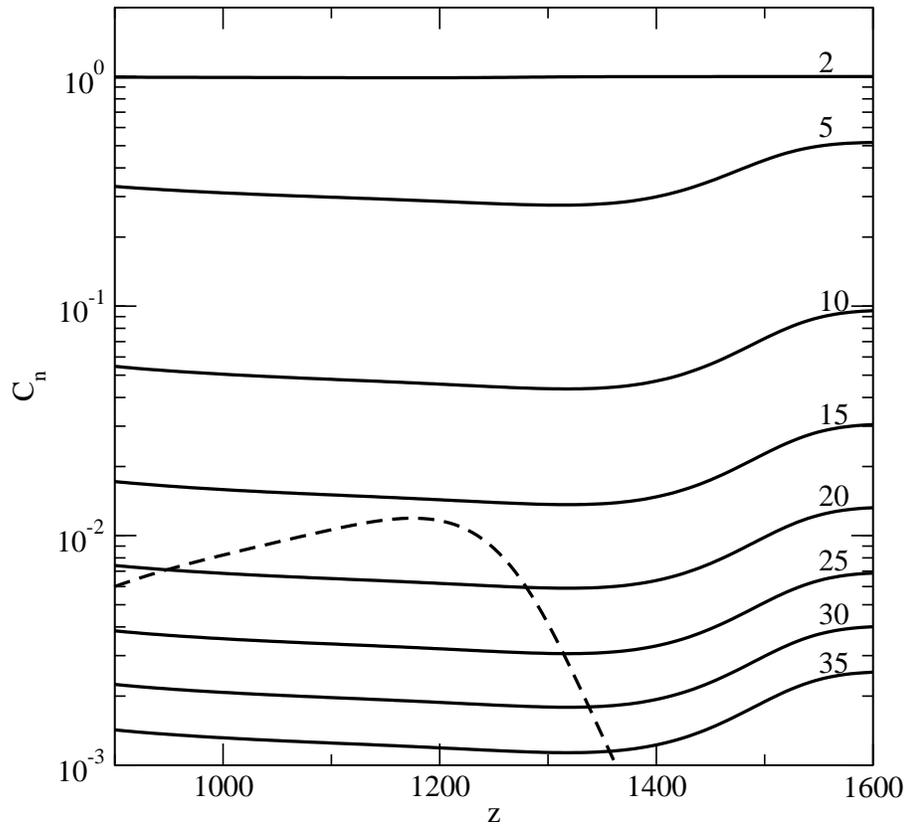}
\end{center}
\caption{
Dependencies of $C_n$ on redshift $z$ are shown by solid lines for 
different values of principal quantum number $n$ indicated near lines. 
Dependence of $\left(1-C_2\right)$ on redshift $z$ is shown by dashed line.
}
\label{fig3}
\end{wrapfigure}

\begin{wrapfigure}[34]{o}{0.9\textwidth}
\begin{center}
\includegraphics[width=0.9\textwidth, clip]{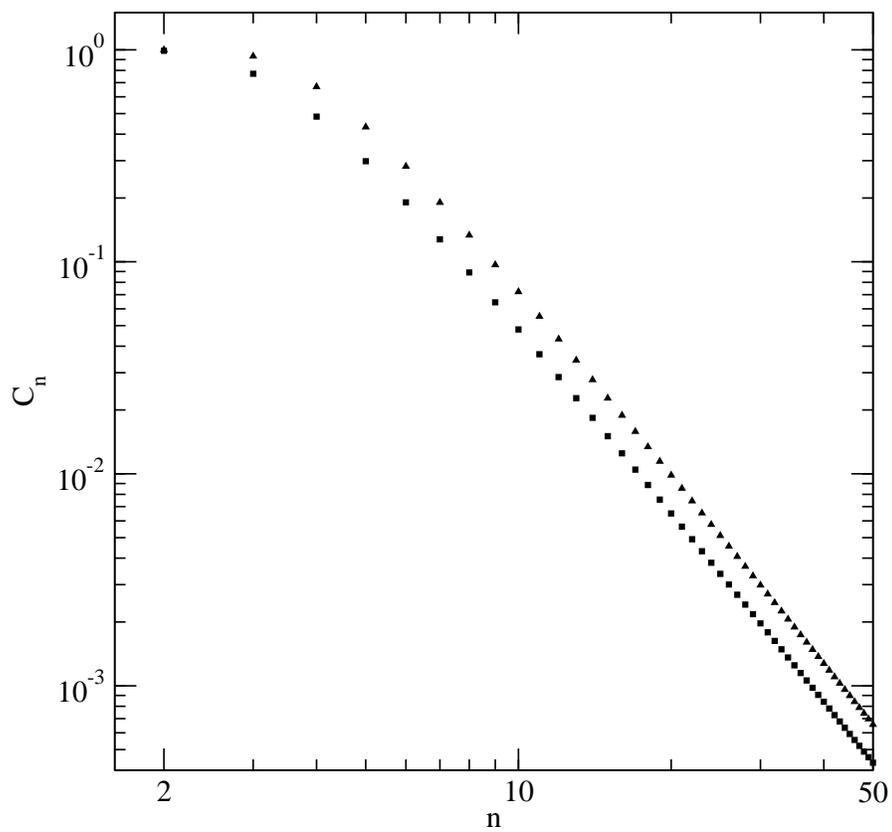}
\end{center}
\caption{
Dependencies of $C_n$ on principal quantum number $n$ are shown 
by triangles (for $z=1500$) and squares (for $z=1100$).
}
\label{fig4}
\end{wrapfigure}

Results of calculations of coefficients $C_n(z)$ are shown in the Figs 
\ref{fig3} and \ref{fig4}. 

%---section devoted to 2s1s transitions, possibly-------------------
%---seems to be non-reasonable because of complexity of stimulated---
%---two-photon transitions with emitting super Ly-alpha photons-----

\subsection{Influence of Feedbacks on Hydrogen Recombination}
Knowledge of coefficients $C_{\rm n}(z)$ allows us to calculate the feedback 
correction to the hydrogen ionization fraction $x_{\rm HII}$ and 
correspondingly to the concentration of free electrons $\Delta N_e/N_e$. 
Ionization fraction $x_{\rm HII}$ is determined by the standard TLA kinetic 
equation (Peebles 1968, Seager et al. 1999) with modified inhibition factor:
\begin{equation}
C_{\rm HI}={A^{\rm r}_{\rm eff} + A_{\rm 2s1s} \over 
\beta_{\rm HI}+A^{\rm r}_{\rm eff} + A_{\rm 2s1s}}
\label{C_HI_res}
\end{equation}
where $A^{\rm r}_{\rm eff}$ is the new total effective coefficient of 
np$\leftrightarrow$1s transitions, $A_{\rm 2s1s}$ is the coefficient of 
2s$\rightarrow$1s two-photon spontaneous transition, 
$\beta_{\rm HI}$ is the effective total ionization coefficient from 
excited states of hydrogen atoms. 
The effective coefficient 
$A^{\rm r}_{\rm eff}$ is given by the following formula:
\begin{equation}
A^{\rm r}_{\rm eff}=\sum_{\rm n\ge 2}C_{\rm n}{g_{\rm n} \over g_{\rm 1}}P_{\rm n}A_{\rm n}
\exp\left(-{E_{\rm n}-E_{\rm 2}\over k_{\rm B}T}\right)
\label{A_r_eff}
\end{equation}
where $E_{\rm n}$ is the energy of level $n$. 
Taking into account the approximated expression for Sobolev escape probability 
\begin{equation}
P_{\rm n}\simeq {g_{\rm 1} 8\pi H \nu_{\rm n}^3 \over g_{\rm n}A_{\rm n} N_{\rm HI} c^3}
\label{A_red_n}
\end{equation}
one can find the correction to the total rate of 
np$\leftrightarrow$1s transitions in comparison with the ``standard'' 
2p$\leftrightarrow$1s rate:
\begin{equation}
{\Delta J \over J_{\rm 2p1s}}=-\left(1-C_{\rm 2}\right)+
\sum_{\rm n\ge 3}C_{\rm n}{\nu_{\rm n}^3 \over \nu_{\rm \alpha}^3}
\exp\left(-{E_{\rm n}-E_{\rm 2}\over k_{\rm B}T}\right)
\label{Delta_A_r}
\end{equation}

Including feedback $3\Rightarrow 2$ leads to the negative 
(i.e. decelerating) correction $\left(1-C_2\right)$ while 
$(n+1)\Rightarrow n$ feedbacks ($n\ge 3$) lead to 
positive (i.e. accelerating) corrections. It has the following 
explanation: coefficients $C_n$ take into account not only re-absorption of 
$(n+1)\rightarrow 1$ photons in $n\rightarrow 1$ line (decelerating part 
of effect which is equal to $\left(C_n-1\right)$) but also $n\rightarrow 1$
transitions due to escape of photons from line profile 
(accelerating part which is equal to $1$). 
Both these effects have not been taken into account in the standard 
TLA-model excepting 2p$\rightarrow$1s transitions due to escape of 
Ly$\alpha$ photons from line profile. 

Note that in traditional notation of recombination kinetic equation 
(e.g. Peebles 1968, Wong et al. 2008) for allowance of hydrogen feedbacks 
one should replace usual Peebles $K_{\rm H}$-factor 
($K^{\rm old}_{\rm H}=c^3/\left(8\pi \nu_{\rm \alpha}^3 H\right)$, where $H$ 
is the Hubble constant as a function of $z$ according formula 
(\ref{Hubble_const})) by the following:
\begin{equation}
K^{\rm new}_{\rm H}=K^{\rm old}_{\rm H}
\left(1+{\Delta J \over J_{\rm 2p1s}}\right)^{\rm -1}
\label{K_new}
\end{equation}

Results of calculations of relative change of free electron 
concentration during hydrogen recombination epoch are shown 
in the Fig. \ref{fig5}. 

For comparison with result by Chluba and Sunyaev (2009) for hydrogen, the 
relative difference of the free electron concentration $\Delta N_e/N_e$ 
between the Feedback TLA model and TLA model taking into account transitions 
from high excited np-states ($n\ge 3$) due to escape of Ly$n$ photons from 
line profiles (e.g. Dubrovich and Grachev 2005) has been calculated 
(see Fig. \ref{fig6}). 
The maximum of this relative difference is 0.259\% at $z\simeq 1055$, 
which is in very good accordance with corresponding result by 
Chluba and Sunyaev (2009).

\subsection{Effect of Neutral Deuterium}
The fact that deuterium Ly$\alpha$ frequency is larger than hydrogen one 
means that redshifted hydrogen Ly$\beta$ photons may be absorbed by neutral 
deuterium earlier than by hydrogen. This may lead to screening of redshifted 
hydrogen Ly$\beta$ radiation from HI 1s$\rightarrow$2p transitions by depth 
of neutral deuterium. Fung and Chluba (2009) have noted that existence 
of such screening depends on concrete conditions of radiation transfer, and 
estimate of screening effect may be very complicated problem in general case. 
In this subsection some necessary details of redshifted HI Ly$\beta$ radiation 
transfer through DI Ly$\alpha$ and HI Ly$\alpha$ lines are considered. 
The absorption coefficients [cm$^{\rm -3}$] for hydrogen and deuterium 
in $2p\rightarrow 1s$ line are given by the following formula
\begin{equation}
\kappa_{\rm X} (\nu, z)={g_{\rm 2p}\over g_{\rm 1s}}A_{\rm X,2p}\phi_{\rm X,2p}(\nu)N_{\rm X,1s}
\label{kappas}
\end{equation} 
where subscript $X$ denotes component ($X=H$ or $D$), 
$\phi_{\rm X,2p}(\nu)$ [Hz$^{\rm -1}$] is the absorption 2p$\rightarrow$1s line 
profile for component $X$ ($\int \phi_{\rm X,2p}\left(\nu\right)d\nu=1$), 
$N_{\rm X,1s}$ is concentration of atoms $X$ in ground state. 

Corresponding optical depths are given by the following formula 
\begin{equation}
\tau_{\rm X} (\nu, z) = \int_{\rm \nu}^{\rm \infty} {c^3 \over 8\pi \nu'^3} 
{\kappa_{\rm X} (\nu', z') \over H(z')}d\nu'
\label{optical_depth}
\end{equation}
where $z'=(1+z){\nu' / \nu}-1$. 

The value $N_{\rm D,1s}$ is controlled mainly by charge transfer reaction 
(Galli and Palla 1998, Stancil et al. 1998 and references therein):
\begin{equation}
D^{+} + H \leftrightarrow D + H^{+}
\label{charge_transfer}
\end{equation}
which tries to keep the fraction of neutral deuterium 
a little larger than the one of neutral hydrogen. This deviation between 
fractions of neutral hydrogen and neutral deuterium is due to difference 
between their ionization energies which in turn is determined by 
isotopic shift for these atoms. This energy difference is 
$\Delta T_{\rm DH}=43$ K in temperature units. Therefore during hydrogen 
recombination epoch $z = 800 - 1600$ the relative 
deviation of neutral deuterium fraction from hydrogen one is an 
order about $\Delta T_{\rm DH}/T\lesssim 0.02 \ll 1$. Neglecting this 
deviation one can obtain the following formula (see also Wolf Savin 2002) 
\begin{equation}
N_{\rm D,1s}\simeq N_{\rm D}{N_{\rm H,1s}\over N_{\rm H}}
\label{N_D1s}
\end{equation}
where $N_{\rm D}$ is the total concentration of deuterium atoms and ions, 
$N_{\rm H}$ is the same for the hydrogen. Using (\ref{N_D1s}) one can 
estimate the absorption coefficient $\kappa_{\rm D}$ and optical depth 
$\tau_{\rm D}$.

Absorption coefficients and optical depths are shown in the Fig \ref{fig7}. 
The difference between central frequencies of absorption coefficients for 
hydrogen and deuterium is due to isotopic shift having relative 
value $m_{\rm e}/\left(2m_{\rm p}\right)\simeq 27.2 \cdot 10^{\rm -5}$. 
The difference between 
magnitudes of absorption coefficients is determined by the natural 
abundance of deuterium relative to hydrogen. The difference between thermal 
(Doppler) widths of absorption coefficients is due to different masses 
for hydrogen and deuterium atoms. 

From top panel of Fig. \ref{fig7} one can see that at the central frequency of 
DI 2p$\rightarrow$1s line the absorption coefficient of deuterium 
is about an order of magnitude larger than hydrogen one (note that for 
hydrogen the wing of profile is at this frequency). 
Nevertheless, despite this one should take into 
account that regions and frequency ranges where the 
interaction between matter and radiation becomes significant are 
determined by the following condition $\tau\sim 1$. From bottom panel of 
Fig. \ref{fig7} one can see that optical depth for deuterium achieves unity at 
the frequency where the optical depth for hydrogen already has the values 
much larger than unity ($\sim 10^4$). 
Therefore, despite the fact that central absorption frequency for deuterium 
is larger than for hydrogen, the actual frequency of interaction for deuterium 
is less than corresponding one for hydrogen. This means that the redshifted  
Ly$\beta$ photons of hydrogen interact with neutral hydrogen at transitions 
HI 1s$\rightarrow$2p earlier than with neutral deuterium at transitions 
DI 1s$\rightarrow$2p, i.e. effect of screening is absent in the considered case.

\section{Helium Feedback}
\subsection{Semi-analytical Consideration of Feedback for Helium}
In the difference from radiation transfer during hydrogen recombination epoch 
when the medium is almost transparent for radiation at non-resonant frequencies 
(i.e. absorption is negligible) the transfer of redshifted HeI n$\rightarrow$1 
photons during helium recombination epoch occurs at the presence of 
sufficient continuum absorption, namely absorption at ionization of 
neutral hydrogen atoms. Interaction of redshifted HeI n$\rightarrow$1 quanta 
with neutral hydrogen leads to the partial thermalization of this 
superequilibrium radiation. 
This should be taken into account in the solution of radiation transfer 
equation, so instead of formula (\ref{eta_n_from_eta_n1}) one should 
use the following:
\begin{equation}
\eta^{+}_{\rm n}(z)=\eta^{-}_{\rm n+1}(z'_{\rm n})\exp\left(-\tau^{\rm H}_{\rm (n+1)\Rightarrow n}
\right)+\eta^{\rm 0}_{\rm n}(z)\left(1-\exp\left(-\tau^{\rm H}_{\rm (n+1)\Rightarrow n}\right)
\right)
\label{eta_n_from_eta_n1_absorp}
\end{equation}
where $\tau^{\rm H}_{\rm (n+1)\Rightarrow n}$ is the optical depth due to 
ionization of neutral hydrogen. 
Since in present work we consider only $2^1P\leftrightarrow 1^1S$ 
$\Rightarrow$ $2^3P\leftrightarrow 1^1S$ feedback for helium recombination 
we can write the following expression for 
$\tau^{\rm H}_{\rm 2^1P\Rightarrow 2^3P}$:
\begin{equation}
\tau^{\rm H}_{\rm 2^1P\Rightarrow 2^3P}(z)=\int_{\rm z}^{\rm z_{\rm 2^3P}}
{c\sigma_{\rm H}
\left(\nu'\right)N_{\rm HI}\left(z'\right) \over H(z')(1+z')}dz'
\end{equation}
where $z_{\rm 2^3P}=\left(\nu_{\rm 2^1P}/\nu_{\rm 2^3P}\right)(1+z)-1$, 
$\sigma_{\rm H}\left(\nu'\right)$ is the ionization crossection of 
hydrogen ground state at frequency $\nu'=\nu_{\rm 2^3P}(1+z')/(1+z)$, 
$N_{\rm HI}\left(z'\right)$ is the 
concentration of neutral hydrogen atoms at the moment with redshift $z'$. 
Using Saha approximation for $N_{\rm HI}$ the optical depth 
$\tau^{\rm H}_{\rm 2^1P\Rightarrow 2^3P}$ can be estimated (with relative 
accuracy within 5\% for $z\ge 1800$) by the following expression:
\begin{equation}
\tau^{\rm H}_{\rm 2^1P\Rightarrow 2^3P}(z)\simeq {c\sigma_{\rm H}\left(\nu_{\rm 2^3P}\right)
N_{\rm HI}\left(z\right) \over H(z)}{k_{\rm B}T\left(z\right)\over I_{\rm H}}
\left(1-\exp\left({I_{\rm H}\over {k_{\rm B}T\left(z_{\rm 2^3P}\right)}}-
{I_{\rm H}\over {k_{\rm B}T\left(z\right)}}\right)\right)
\label{tau_H_21P_23P}
\end{equation}
where $I_{\rm H}$ is the ionization energy of hydrogen ground state. 

Likewise the hydrogen case, let us define 
$\Gamma_{\rm He}=\left(\eta_{\rm 2^1P}/\eta^{\rm 0}_{\rm 2^1P}\right)-1$, where 
$\eta_{\rm 2^1P}$ is the actual occupation number in 
HeI $2^1P\rightarrow 1^1S$ line, 
$\eta^{\rm 0}_{\rm 2^1P}$ is the equilibrium occupation number in 
HeI $2^1P\rightarrow 1^1S$ line.

Substitution of (\ref{eta_n_from_eta_n1_absorp}) into (\ref{C_n_eta}) 
gives us the following formula:
\begin{equation}
C_{\rm 2^3P}=1-{\Gamma_{\rm He}(z_{\rm 2^3P}) \over \Gamma_{\rm He}(z)}
\exp\left(-\tau^{\rm H}_{\rm 2^1P\Rightarrow 2^3P}\right)
\label{C_n_main_He}
\end{equation}

Dependencies of $C_{\rm 2^3P}$ and $\tau^{\rm H}_{\rm 2^1P\Rightarrow 2^3P}$ on redshift 
$z$ are shown in the Fig. \ref{fig8}. 
%--------------------------------------------------------

\subsection{Influence of Feedback on Helium Recombination}
Knowledge of coefficient $C_{\rm 2^3P}(z)$ allows us to calculate the feedback 
correction to the helium ionization fraction 
$x_{\rm HeII}=N_{\rm HeII}/N_{\rm He}$ and 
correspondingly to the concentration of free electrons $\Delta N_e/N_e$. 
Ionization fraction is determined by the standard TLA kinetic equation 
(Matsuda et al. 1969, Seager et al. 1999, Kholupenko et al. 2007, 
Wong et al. 2008) with modified inhibition factor for ortho-channel 
of recombination:
\begin{equation}
C_{\rm or}={C_{\rm 2^3P}\left(g_{\rm 2^3P} / g_{\rm 2^3S} \right)
A_{\rm 2^3P}\left(P^{\rm H}_{\rm 2^3P}+P^{\rm r}_{\rm 2^3P}\right)
\exp\left(-{E_{\rm 2^3P2^3S}/ k_BT}\right) 
\over \beta_{\rm or}+C_{\rm 2^3P}\left(g_{\rm 2^3P} / g_{\rm 2^3S} \right)
A_{\rm 2^3P}\left(P^{\rm H}_{\rm 2^3P}+P^{\rm r}_{\rm 2^3P}\right)
\exp\left(-{E_{\rm 2^3P2^3S}/ k_BT}\right)}
\label{C_HeI_res}
\end{equation}
where $A_{\rm 2^3P}$ is the coefficient of HeI $2^3P\rightarrow 1^1S$ 
spontaneous transition, $E_{\rm 2^3P2^3S}$ is the $2^3P\rightarrow 2^3S$ 
transition energy, $\beta_{\rm or}$ is the effective total ionization 
coefficient of helium ortho-states, 
$P^{\rm H}_{\rm 2^3P}$ is the probability of HeI $2^3P\rightarrow 1^1S$ 
transitions due to destruction of HeI $2^3P\rightarrow 1^1S$ resonant photons 
by neutral hydrogen within line (Switzer and Hirata 2008, Kholupenko et al. 2007, 
Rubino-Martin et al. 2008), $P^{\rm r}_{\rm 2^3P}$ is the modified Sobolev escape 
probability of HeI $2^3P\rightarrow 1^1S$ transitions (Kholupenko et al. 2008). 
Expressions for probabilities $P^{\rm H}_{\rm 2^3P}$ and $P^{\rm r}_{\rm 2^3P}$  appropriate 
for TLA-models can be found in Wong et al. (2008) and Kholupenko et al. (2008). 
For comparison with Chluba and Sunyaev (2009), the dependence of 
the effective escape probability 
$C_{\rm 2^3P}\left(P^{\rm H}_{\rm 2^3P}+P^{\rm r}_{\rm 2^3P}\right)$ 
on redshift $z$ is presented in Fig. \ref{fig9}. This function 
is in very good accordance with result by Chluba and Sunyaev (2009). 

Dependencies of the free electron fraction normalized by total concentration 
of hydrogen $N_e/N_H$ and correction $\Delta N_e/N_e$ on redshift $z$ for the 
helium recombination epoch are presented in Figs \ref{fig10} 
and \ref{fig11} correspondingly. 

In the Fig. \ref{fig11} for additional comparison with 
Chluba and Sunyaev (2009), $\Delta N_e/N_e$ in the case of 
partial taking into account hydrogen continuum absorption is presented 
(``partial'' means that radiation transfer within the resonant 
$2^1P\leftrightarrow 1^1S$ and 
$2^3P\leftrightarrow 1^1S$ lines has been calculated with taking into account 
the presence of neutral hydrogen, but radiation transfer between these lines 
has been calculated without the presence of neutral hydrogen). 
In this case the maximum of $\Delta N_e/N_e$ is 0.372\% at $z\simeq 2018$. 
Such perturbation of free electron number 
can affect the visibility function much larger than in the case of 
complete taking into account hydrogen continuum absorption (i.e. that would 
be important for analysis of damping tail of CMB anisotropy). 
Obtained result is in very good accordance with corresponding one by 
Chluba and Sunyaev (2009). Once again it should be noted that result 
for partial taking into account hydrogen continuum absorption is not 
valid and was obtained as intermediate one for comparison only. 

\section{Results}
The main result of this paper is the semi-analytical approach 
(formulae \ref{Gamma_H}, \ref{C_n_main}, \ref{C_HI_res} - \ref{K_new}, 
\ref{tau_H_21P_23P} - \ref{C_HeI_res}) describing the feedback effect 
for resonant transitions in cosmological plasma. This approach allows 
us to calculate the free electron concentration $N_{e}$ taking 
into account feedback effect and correction $\Delta N_e/N_e$ 
relative to the ``standard'' free electron concentration obtained 
in the frame of common TLA model 
(i.e. TLA model without taking into account transitions from high 
excited states $n\ge 3$ and feedbacks, e.g. {\bf recfast} by Wong et 
al. 2008).  

\begin{wrapfigure}[34]{o}{0.9\textwidth}
\begin{center}
\includegraphics[width=0.9\textwidth, clip]{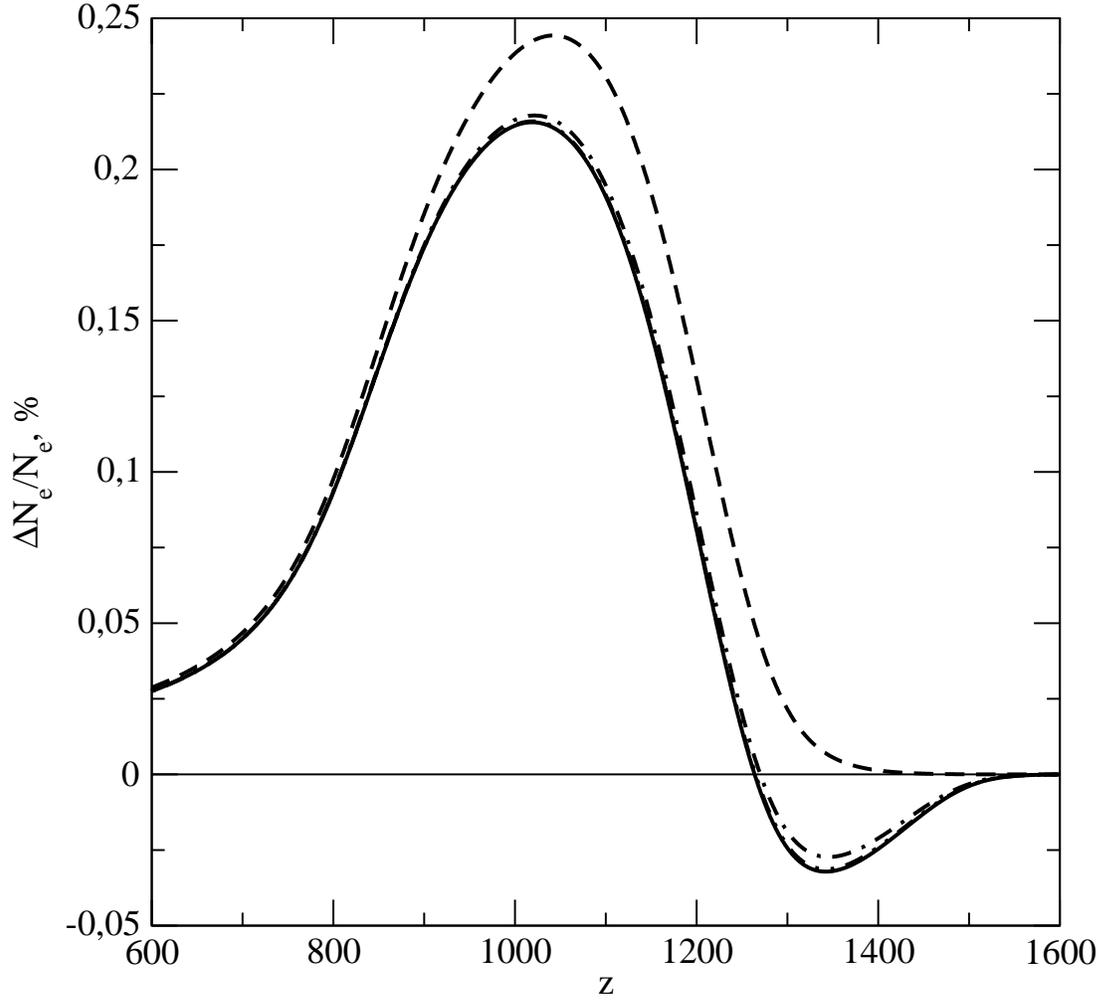}
\end{center}
\caption{
Convergence of results at variation of number of levels 
involved in calculation. Curves presented are the relative 
differences between Feedback TLA model and common TLA model 
(without transitions from high excited states ($n\ge 3$) and feedbacks): 
result for $n_{\rm max}=2$ (i.e. feedback $3\Rightarrow 2$ is taken into 
account) is shown by dashed line, 
result for $n_{\rm max}=3$ (i.e. feedbacks $4\Rightarrow 3\Rightarrow 2$ are 
taken into account) is shown by dashed-dotted line,
result for $n_{\rm max}=5$ (i.e. feedbacks $6\Rightarrow ... \Rightarrow 2$ are 
taken into account) is shown by dashed-dotted line,
result for $n_{\rm max}=30$ (i.e. feedbacks $31\Rightarrow ... \Rightarrow 2$ are 
taken into account) is shown by solid line. Note that curve for $n_{\rm max}=5$ 
is very close to the curve for $n_{\rm max}=30$.
}
\label{fig5}
\end{wrapfigure}

\begin{wrapfigure}[34]{o}{0.9\textwidth}
\begin{center}
\includegraphics[width=0.9\textwidth, clip]{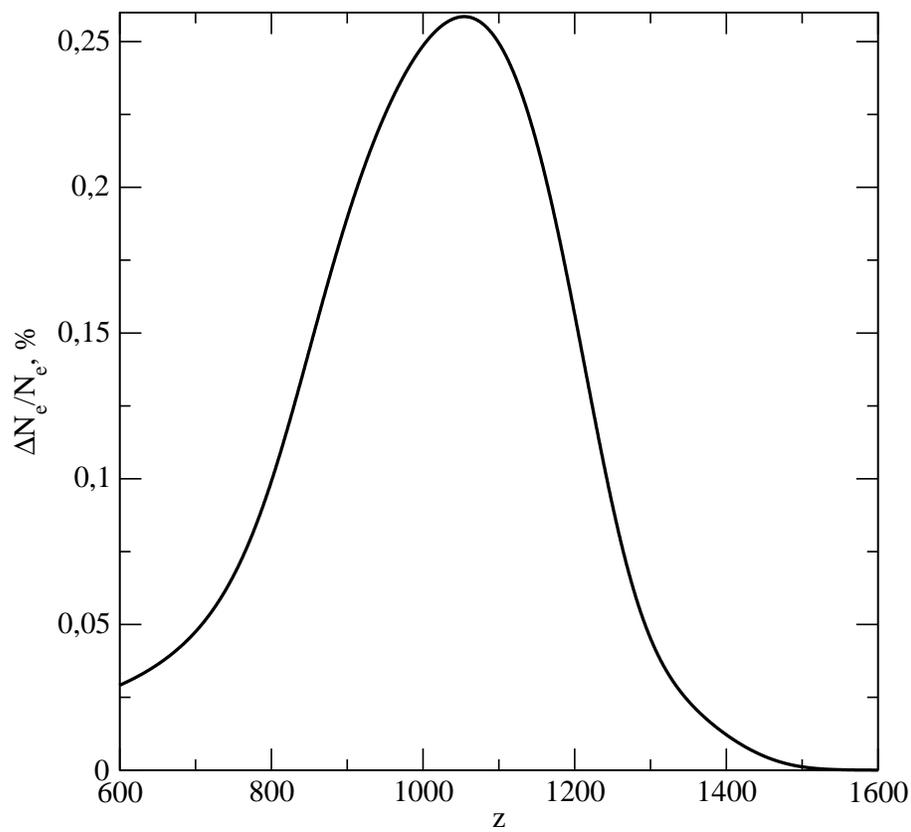}
\end{center}
\caption{
The relative difference of free electron concentration 
between Feedback TLA model and TLA model 
taking into account transitions from high excited np-states $n\ge 3$ 
(i.e. TLA model with $C_{\rm n}=1$). This result is 
obtained for the number of levels $n_{\rm max}=10$.
}
\label{fig6}
\end{wrapfigure}

\begin{wrapfigure}[34]{o}{0.9\textwidth}
\begin{center}
\includegraphics[width=0.9\textwidth, clip]{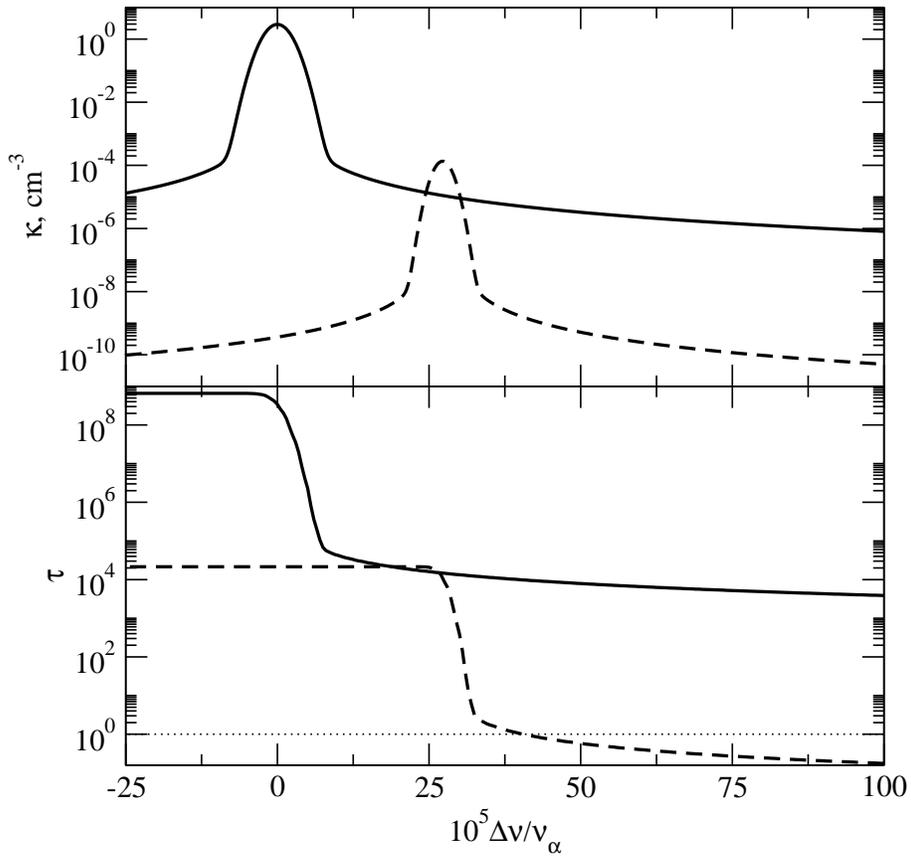}
\end{center}
\caption{
{\bf Top panel:} 1s$\rightarrow$2p absorption coefficients for hydrogen 
(solid line) and deuterium (dashed line) as functions of relative frequency 
deviation from central frequency of hydrogen Ly$\alpha$-line. 
Curves are shown for moment $z=1300$. 
{\bf Bottom panel:} optical depths of absorption at HI 1s$\rightarrow$2p 
transition (solid line) and at DI 1s$\rightarrow$2p transition (dashed line). 
Dotted line shows the unity level.
}
\label{fig7}
\end{wrapfigure}

\begin{wrapfigure}[34]{o}{0.9\textwidth}
\begin{center}
\includegraphics[width=0.9\textwidth, clip]{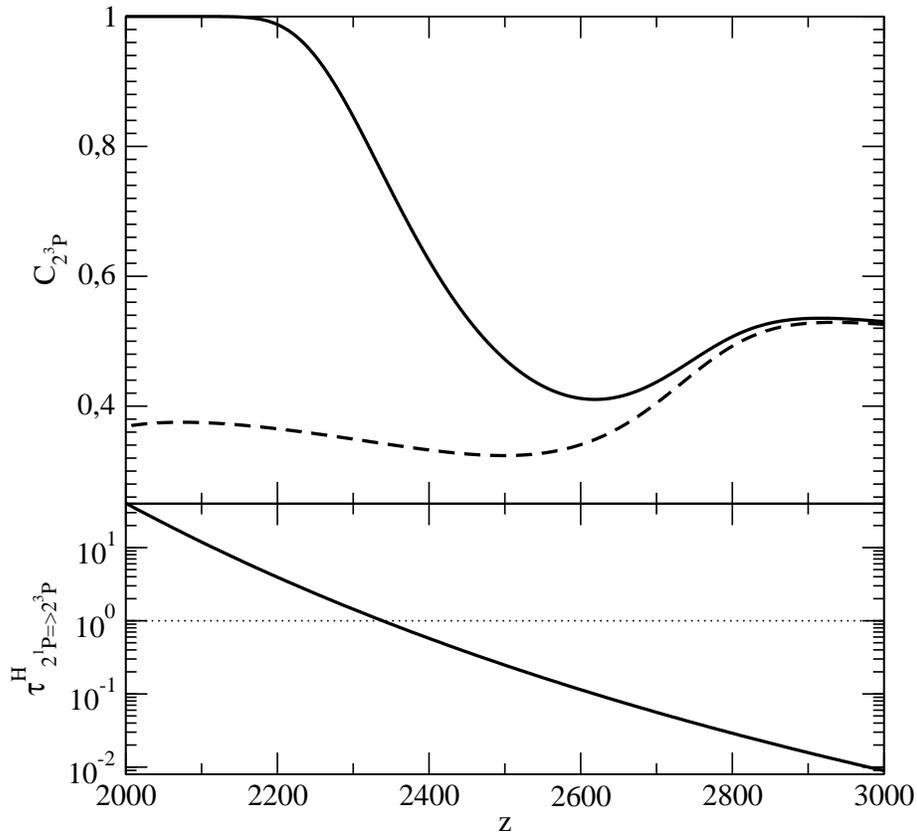}
\end{center}
\caption{
{\bf Top panel:} Coefficient $C_{\rm 2^3P}$ as a function of redshift $z$ 
with taking into account hydrogen continuum absorption (solid line) 
and without taking into account hydrogen continuum absorption (i.e. 
for $\tau^{\rm H}_{\rm 2^1P\Rightarrow 2^3P}=0$, dashed line)
{\bf Bottom panel:} Optical depth $\tau^{\rm H}_{\rm 2^1P\Rightarrow 2^3P}$ 
as a function of redshift $z$ (according formula (\ref{tau_H_21P_23P})). 
Dotted line shows the unity level.
}
\label{fig8}
\end{wrapfigure}

\begin{wrapfigure}[34]{o}{0.9\textwidth}
\begin{center}
\includegraphics[width=0.9\textwidth, clip]{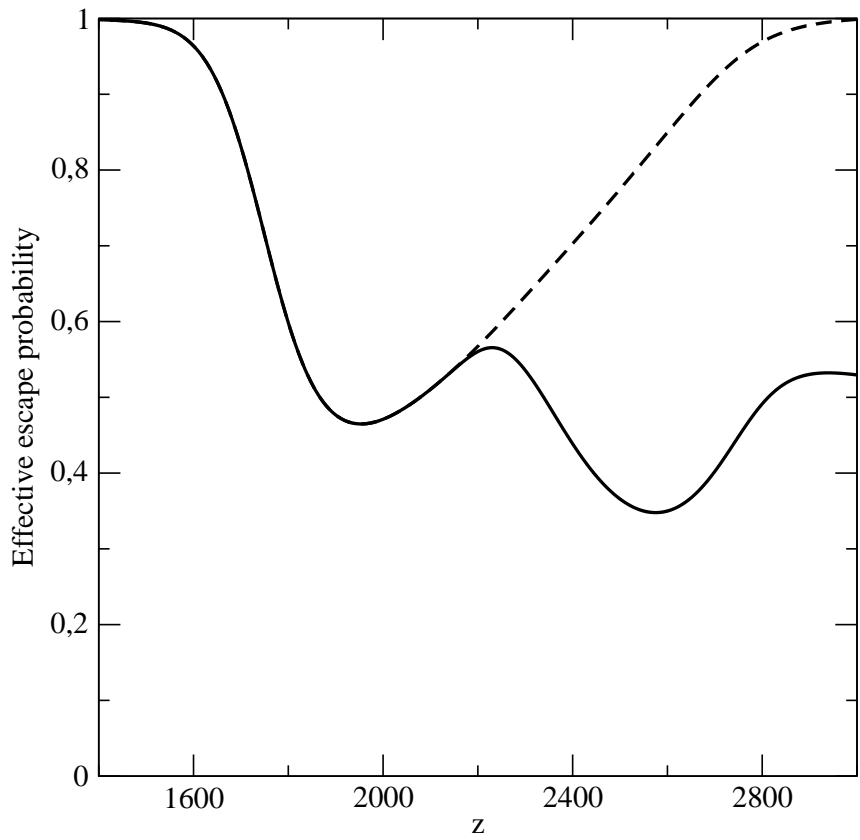}
\end{center}
\caption{Effective escape probability 
$C_{\rm 2^3P}\left(P^{\rm H}_{\rm 2^3P}+P^{\rm r}_{\rm 2^3P}\right)$ 
as a function of redshift $z$ with taking into account 
feedback (solid line) and without (i.e. for $C_{\rm 2^3P}=1$, dashed line)
}
\label{fig9}
\end{wrapfigure}

\begin{wrapfigure}[34]{o}{0.9\textwidth}
\begin{center}
\includegraphics[width=0.9\textwidth, clip]{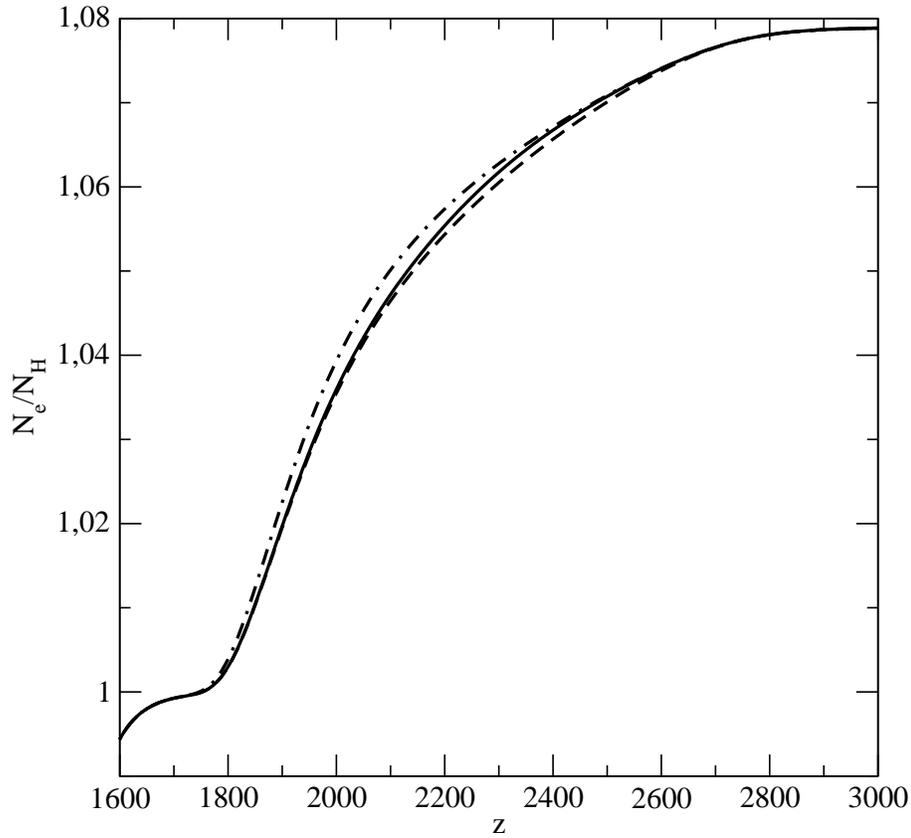}
\end{center}
\caption{Dependence of $N_e/N_H$ (relative number of free electron 
normalized by the total number of hydrogen atoms and ions) on 
redshift $z$: the result for model without feedback is shown by 
dashed line, the result for model with feedback and partial taking 
into account hydrogen continuum absorption (i.e. $P^{\rm H}_{\rm 2^1P}$ 
and $P^{\rm H}_{\rm 2^3P}$ take into account hydrogen continuum absorption, 
but $\tau^{\rm H}_{\rm 2^1P\Rightarrow 2^3P}=0$) 
is shown by dashed-dotted line, 
the result for model with feedback and complete taking into account 
hydrogen continuum absorption ($\tau^{\rm H}_{\rm 2^1P\Rightarrow 2^3P}$ is 
calculated by formula (\ref{tau_H_21P_23P})) is shown by solid line.
}
\label{fig10}
\end{wrapfigure}

The correction to the free electron concentration $\Delta N_e/N_e$ 
during hydrogen recombination epoch is presented in the Fig. \ref{fig5}. 
The maximal value of this correction is 0.216\% at redshift about 
$z\simeq 1020$. It can lead to the anisotropy power spectrum correction 
having the order about 0.1 - 0.2\%. 

\begin{wrapfigure}[34]{o}{0.9\textwidth}
\begin{center}
\includegraphics[width=0.9\textwidth, clip]{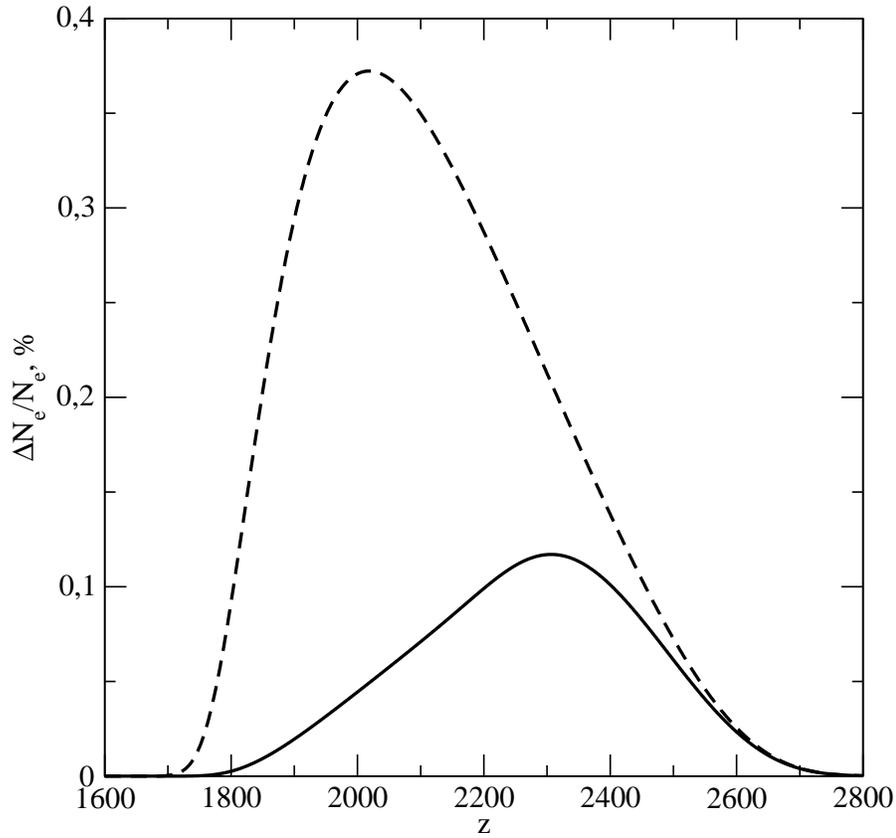}
\end{center}
\caption{
Relative change of free electron concentration $\Delta N_e/N_e$ as a 
function of redshift $z$: (i) 
the result for model with feedback and partial taking 
into account hydrogen continuum absorption (i.e. $P^{\rm H}_{\rm 2^1P}$ 
and $P^{\rm H}_{\rm 2^3P}$ take into account hydrogen continuum absorption, 
but $\tau^{\rm H}_{\rm 2^1P\Rightarrow 2^3P}=0$) is shown by dashed line, 
(ii) the result for model with feedback and complete taking into account 
hydrogen continuum absorption ($\tau^{\rm H}_{\rm 2^1P\Rightarrow 2^3P}$ is 
calculated by formula (\ref{tau_H_21P_23P})) is shown by solid line.
}
\label{fig11}
\end{wrapfigure}

The correction to the free electron concentration $\Delta N_e/N_e$ 
during helium recombination epoch is presented in the Fig. \ref{fig11}. 
The maximal value of this correction is 0.117\% at redshift about 
$z\simeq 2308$. 
%At redshift $z=2000$ the correction $\Delta N_e/N_e$ is less than 0.05\%. 
Such value of correction is negligible at the current level of experimental 
accuracy (e.g. needed for treatment of Planck experimental data). 

All results obtained in this work (in the frame of modified TLA model) 
are in very good accordance with corresponding results 
obtained in the frame of multilevel model by Chluba and Sunyaev (2009).

The main conclusions of present work are the following: 
\\1) The modification of ionization history of the Universe due to 
HI Ly$(n+1)$ $\Rightarrow$  HI n$\leftrightarrow$1 feedbacks is 
important for correct analysis of CMB anisotropy power spectrum 
at the current level of experimental accuracy (e.g. Planck mission). 
This modification in particular can be taken into account by using 
semi-analytical approach developed in present paper. 
\\2) The modification of ionization history of the Universe due to 
HeI $2^1P\leftrightarrow 1^1S$ $\Rightarrow$ $2^3P\leftrightarrow 1^1S$ 
feedback 
is negligible for correct analysis of CMB anisotropy power spectrum 
at the current level of experimental accuracy. The smallness of this 
modification is particularly provided by screening of 
HeI $2^1P\leftrightarrow 1^1S$ resonant quanta from HeI 
$2^3P\leftrightarrow 1^1S$ transition by small amount of neutral 
hydrogen existing during helium recombination. It allows 
us to avoid accounting the feedbacks for helium at modeling 
of cosmological recombination. 

\vspace{0.3cm}
\hspace{-1.4cm}
{\bf Acknowledgements}
\\Authors thank M.S. Burgin for pointing out 
the problem of possible screening of HI Ly$\beta$ radiation by neutral 
deuterium. Also authors thank J. Chluba and J. Fung who have 
clarified the important role of optical depth for deuterium screening. 
Authors are grateful to participants of Workshops ``Physics of 
Cosmological Recombination'' (Max Planck Institute for Astrophysics, 
Garching 2008 and University Paris-Sud XI, Orsay 2009) for 
useful discussions on cosmological recombination. 
%Special thanks to V.M. Bessolova who provides our works during many years.

This work has been performed with support of RFBR grant 08-02-01246a, 
grant "Leading Scientific Schools of Russia" NSh-2006.2008.2 and grant of 
Dynasty Foundation.

\vspace{0.3cm}
\hspace{-1.4cm}
{\bf References}
\\Burgin M.S., Astronomy Reports, 47, 709 (2003)
\\Chluba J., Sunyaev R.A., A\&A, 475, 109 (2007)
\\Chluba J., Sunyaev R.A., arXiv:0909.2378 (2009)
\\Dubrovich V.K., Grachev S.I., Astronomy Letters, 31, 359 (2005)
\\Fendt W.A., Chluba J., Rubino-Martin J.A., Wandelt B.D., 
Ap. J. Supplement, 181, 627 (2009).
\\Fung J., Chluba J., in preparation (2009)
\\Galli D., Palla F., A\&A, 335, 403 (1998)
\\Grachev S.I., Dubrovich V.K., ASTROPHYSICS, 34, 124 (1991)
\\Grin D., Hirata C.M., arXiv:0911.1359
\\Kholupenko E.E., Ivanchik A.V., Astronomy Letters, 32, 795 (2006)
\\Kholupenko E.E., Ivanchik A.V., Varshalovich D.A., MNRAS Letters, 
378, L39 (2007)
\\Kholupenko E.E., Ivanchik A.V., Varshalovich D.A., Astronomy Letters, 
34, 725 (2008)
\\Matsuda T., Sato H., Takeda H., Progr. Theor. Physics, 42, 219 (1969)
\\Peebles P. J., ApJ, 153, 1 (1968)
\\Rubino-Martin J.A., Chluba J., Sunyaev R.A., MNRAS, 371, 1939 (2006)
\\Rubino-Martin J.A., Chluba J., Sunyaev R.A., A\&A, 485, 377 (2008) 
\\Seager S., Sasselov D.D., Scott D., ApJ, 523, L1 (1999)
\\Stancil P.C., Lepp S., Dalgarno A., ApJ, 509, 1 (1998)
\\Switzer E.R., Hirata C.M., Phys.Rev. D, 77, id. 083006 (2008)
\\Wolf Savin D., ApJ, 566, 599 (2002)
\\Wong W.Y., Moss A., Scott D., MNRAS, 386, 1023 (2008)
\\Zeldovich Y.B., Kurt V.G., Sunyaev R. A., ZhETF, 55, 278 (1968) 

%\\Chluba J., Rubino-Martin J.A., Sunyaev R.A., MNRAS, 374, 1310 (2007)
%\\Hirata C.M., Switzer E.R., Phys.Rev. D, 77, id. 083007 (2008)
%\\Switzer E.R., Hirata C.M., Phys.Rev. D, 77, id. 083008 (2008)

%%%%%%%%%%%%%%%%%%%%%%%%%%% FIGURE 1 %%%%%%%%%%%%%%%%%%%%%%%%%%%%%
\newpage

%%%%%%%%%%%%%%%%%%%%%%%%%%% FIGURE 2 %%%%%%%%%%%%%%%%%%%%%%%%%%%%%
\newpage

%%%%%%%%%%%%%%%%%%%%%%%%%%% FIGURE 3 %%%%%%%%%%%%%%%%%%%%%%%%%%%%%
\newpage

%%%%%%%%%%%%%%%%%%%%%%%%%%% FIGURE 4 %%%%%%%%%%%%%%%%%%%%%%%%%%%%%
\newpage

%%%%%%%%%%%%%%%%%%%%%%%%%%% FIGURE 5 %%%%%%%%%%%%%%%%%%%%%%%%%%%%%
\newpage

%%%%%%%%%%%%%%%%%%%%%%%%%%% FIGURE 6 %%%%%%%%%%%%%%%%%%%%%%%%%%%%%
\newpage

%%%%%%%%%%%%%%%%%%%%%%%%%%% FIGURE 7 %%%%%%%%%%%%%%%%%%%%%%%%%%%%%
\newpage

%%%%%%%%%%%%%%%%%%%%%%%%%%% FIGURE 8 %%%%%%%%%%%%%%%%%%%%%%%%%%%%%
\newpage

%%%%%%%%%%%%%%%%%%%%%%%%%%% FIGURE 9 %%%%%%%%%%%%%%%%%%%%%%%%%%%%%
\newpage

%%%%%%%%%%%%%%%%%%%%%%%%%%% FIGURE 10 %%%%%%%%%%%%%%%%%%%%%%%%%%%%%
\newpage

%%%%%%%%%%%%%%%%%%%%%%%%%%% FIGURE 11 %%%%%%%%%%%%%%%%%%%%%%%%%%%%%
\newpage
\end{document}